\documentclass{article}

\input{tcilatex}

\begin{document}

\begin{center}
{\LARGE NEW SPIN\ SQUEEZING\ AND\bigskip }

{\LARGE OTHER\ ENTANGLEMENT\ TESTS\ FOR\bigskip\ }

{\LARGE TWO MODE SYSTEMS\ OF\ \bigskip }

{\LARGE IDENTICAL BOSONS\ \ \bigskip }

B. J. Dalton$^{1,2}$, L. Heaney$^{3}$, J. Goold$^{1,4}$, B. M. Garraway$^{5}$
and Th. Busch$^{1,6}${\LARGE \bigskip }

$^{1}$\textit{Physics Department, University College Cork, Cork City, Ireland%
}

$^{2}$\textit{Centre for Atom Optics and Ultrafast Spectroscopy, Swinburne
University of Technology, Melbourne, Victoria 3122, Australia}

$^{3}$\textit{Centre for Quantum Technologies, National University of
Singapore, Singapore 117543.}

$^{4}$\textit{Clarendon Laboratory, Oxford University, Oxford OX1 3PU,
United Kingdom}

$^{5}$\textit{Department of Physics and Astronomy, University of Sussex,
Falmer, Brighton BN1 9QH, United Kingdom}

$^{6}$\textit{Quantum Systems Unit, Okinawa Institute of Science and
Technology Graduate University, Okinawa, Japan\bigskip }

E-mail: bdalton@swin.edu.au \textit{\bigskip }
\end{center}

\textbf{Abstract \qquad }For any quantum state representing a physical
system of identical particles, the density operator must satisfy the
symmetrisation principle (SP) and conform to super-selection rules (SSR)
that prohibit coherences between differing total particle numbers. Here we
consider bi-partitite states for massive bosons, where both the system and
subsystems are modes (or sets of modes) and particle numbers for quantum
states are determined from the mode occupancies. Defining non-entangled or
separable states as those prepared via LOCC processes, the subsystem density
operators are also required to satisfy the SP and conform to the SSR, in
contrast to some other approaches. Whilst in the presence of this additional
constraint the previously obtained sufficiency criteria for entanglement,
such as the sum of the $\widehat{S}_{x}$ and $\widehat{S}_{y}$ variances for
the Schwinger spin components being less than half the mean boson number,
and the strong correlation test of $|\left\langle \widehat{a}^{m}\,(\widehat{%
b}^{\dag })^{n}\right\rangle |^{2}$ being greater than $\left\langle (%
\widehat{a}^{\dag })^{m}\widehat{a}^{m}\,(\widehat{b}^{\dag })^{n}\widehat{b}%
^{n}\right\rangle (m,n=1,2,..)$ are still valid, new tests are obtained in
our work. We show that the presence of spin squeezing in at least one of the
spin components $\widehat{S}_{x}$, $\widehat{S}_{y}$ and $\widehat{S}_{z}$
is a sufficient criterion for the presence of entanglement and a simple
correlation test can be constructed of $|\left\langle \widehat{a}^{m}\,(%
\widehat{b}^{\dag })^{n}\right\rangle |^{2}$ merely being greater than zero.
We show that for the case of relative phase eigenstates, the new spin
squeezing test for entanglement is satisfied (for the principle spin
operators), whilst the test involving the sum of the $\widehat{S}_{x}$ and $%
\widehat{S}_{y}$ variances is not. However, another spin squeezing
entanglement test for Bose-Einstein condensates involving the variance in $%
\widehat{S}_{z}$ being less than the sum of the squared mean values for $%
\widehat{S}_{x}$ and $\widehat{S}_{y}$ divided by the boson number was based
on a concept of entanglement inconsistent with the SP, and here we present a
revised treatment which again leads to spin squeezing as an entanglement
test.\bigskip

$^{2}$Author to whom any correspondence may be addressed.\bigskip

\textbf{Contents\bigskip }

1. Introduction\qquad \qquad \qquad \qquad \qquad \qquad \qquad \qquad
\qquad \qquad \qquad \qquad \qquad 2

2. Global and local SSR\qquad \qquad \qquad \qquad \qquad \qquad \qquad
\qquad \qquad \qquad \qquad 4

3. Spin squeezing requires entanglement\qquad \qquad \qquad \qquad \qquad
\qquad \qquad\ \ \ \ \ 6

4. Other spin squeezing tests and local SSR separable states\qquad \qquad
\qquad\ \ 6

\qquad 4.1 Hillery et al 2006 spin squeezing test\qquad \qquad \qquad \qquad
\qquad \qquad\ \ \ \ 7

\qquad 4.2 Sorensen et al 2001 spin squeezing test\qquad \qquad \qquad
\qquad \qquad \qquad 8

5. Non spin squeezing tests and local SSR separable states\qquad \qquad
\qquad\ \ \ \ 9

6. Experimental considerations\qquad \qquad \qquad \qquad \qquad \qquad
\qquad \qquad \qquad\ \ \ 10

7. Conclusions\qquad \qquad \qquad \qquad \qquad \qquad \qquad \qquad \qquad
\qquad \qquad \qquad\ \ \ \ \ \ 10

Acknowledgements\qquad \qquad \qquad \qquad \qquad \qquad \qquad \qquad
\qquad \qquad \qquad\ \ \ \ \ 10

References\bigskip \qquad \qquad \qquad \qquad \qquad \qquad \qquad \qquad
\qquad \qquad \qquad \qquad \qquad\ \ \ \ \ 10

\textbf{1. Introduction\bigskip }

Since the work of Einstein et al \cite{epr} on local realism, the famous cat
paradox of Schrodinger \cite{schro} and the derivation of inequalites by
Bell \cite{bell} and others \cite{clauser}, entanglement has been recognised
as being one of the essential features that distinguishes quantum physics
from classical physics. In macroscopic systems entanglement is of\textbf{\ }%
particular importance as it blurs the boundary between the classical and
quantum worlds. This paper considers tests that are sufficient (though not
necessary) to confirm such entanglement.

One way to detect macroscopic entanglement is by applying the so called spin
squeezing inequalities \cite{Wodkiewicz85a} to a large number of particles.
The importance of spin squeezing in quantum metrology was emphasised by
Kitagawa and Ueda \cite{Kitagawa:93} and it has been demonstrated that such
states beat the standard quantum limit in interferometry \cite{Wineland:94}.
For spin angular momentum operators $\hat{S}_{x},\hat{S}_{y},\hat{S}_{z}$
spin squeezing of $\hat{S}_{x}$ with respect to $\hat{S}_{y}$ is defined as 
\begin{equation}
\langle \Delta \hat{S}_{x}^{2}\rangle <\frac{1}{2}|\langle \hat{S}%
_{z}\rangle |\quad \text{with}\quad \langle \Delta \hat{S}_{y}^{2}\rangle >%
\frac{1}{2}|\langle \hat{S}_{z}\rangle |,  \label{eq:spinsq1}
\end{equation}%
where $\langle \Delta \hat{S}_{\eta }^{2}\rangle $ is the variance of the
spin operator $\hat{S}_{\eta }$ ($\eta =x,y,z$) \ Analogous definitions
apply to other pairs of spin operators, or to a spin operator and any of its
perpendicular components. For $N$ spin-1/2 distinguishable particles each in
the same coherent state, correlations between the spins resulted in spin
squeezing in the total spin components \cite{Kitagawa:93}, as it allows the
fluctuations in one direction perpendicular to the spin component to be
reduced. In such systems of distinguishable particles Sorensen et al \cite%
{Sorensen:01, Sorensen:01a} showed that spin squeezing required entanglement
of the $N$\ particles. Here however, we consider $N$\ identical bosons
occupying two modes, where the Schwinger angular momentum operators for two
modes, $A$ and $B$ with annihilation operators $\widehat{a},\widehat{b}$ are 
$\hat{S}_{x}=(\hat{b}^{\dag }\hat{a}+\hat{a}^{\dag }\hat{b})/2,\,\hat{S}%
_{y}=(\hat{b}^{\dag }\hat{a}-\hat{a}^{\dag }\hat{b})/2i$ and $\hat{S}_{z}=(%
\hat{b}^{\dag }\hat{b}-\hat{a}^{\dag }\hat{a})/2$ and the system behaves
like a giant spin with spin quantum number $S=N/2$.

A notable area in which the detection of macroscopic entanglement can take
place is that of trapped ultra-cold gases, where entanglement may be present
over micron scales. Moreover, since non-linear interactions between
particles are required to generate spin squeezing \cite{Kitagawa:93} 
, ultra-cold gases are an ideal test bed for witnesses based on
spin-squeezing inequalities due to the tunable atom-atom interaction. In
addition to achieving interferometry beyond the standard quantum limit,
existing experiments have claimed to demonstrate macroscopic entanglement in
ultra-cold gases via witnesses based on spin squeezing inequalities \cite%
{Esteve:08, experiments, Gross:10}.

However, when detecting entanglement in systems of identical massive bosons
with such inequalities care must be taken for two reasons. Firstly,
entanglement 
is a relative concept that crucially depends on which sub-systems are being
considered. A quantum state may be entangled for one choice of the
sub-systems but may be non-entangled if another choice of sub-systems is
made. For distinguishable particles, the subsystems are usually individual
particles, with internal (spin, polarization) or external (position,
momentum) degrees of freedom possessing the quantum correlations. However,
for identical particles the individual atoms cannot be distinguishable
subsystems, and entanglement will refer to the quantum field modes that the
particles may occupy \cite{Simon:02}. The modes are orthonormal single
particle states, which may be localised in different spatial regions or may
be delocalised where the opposite applies. The system and sub-systems are
now modes, and cases with differing $N$\ are just different quantum states
of the same system. Note however that a different concept of entanglement -
particle entanglement - has also been applied to identical particle systems 
\cite{Wiseman:03a, Hyllus:12a}. Here we formulate spin squeezing
entanglement witnesses for massive identical bosons as above in Eq.(\ref%
{eq:spinsq1}) via second quantization, so the symmetrization principle
applies to all the identical bosons.

In the present paper we follow the approach of Werner \cite{Werner:89a} and
define separable states as those which can be prepared via local operations
(on the sub-systems) and classical communication processes (LOCC). A
relevant example of such states occurs in ultra-cold atom experiments, where
mesoscopic ensembles can be prepared with definite particle numbers in
optical lattice sites. This means that for bipartite systems the density
operator for separable states can be written in the form

\begin{equation}
\hat{\rho}_{sep}=\sum_{R}P_{R}\,\hat{\rho}_{R}=\sum_{R}P_{R}\,\hat{\rho}%
_{R}^{A}\otimes \hat{\rho}_{R}^{B}  \label{eq:separable}
\end{equation}%
where sub-systems $A$, $B$ have been prepared in correlated quantum states $%
\widehat{\rho }_{R}^{A}$ and $\widehat{\rho }^{B}$ with a probability $P_{R}$%
. Entangled states are those which are not of this form. This definitition
can be straightforwardly extended to multiple modes. It is important to
realise that the terms in $\hat{\rho}_{sep}$ have a physical meaning - the
sub-system quantum states $\widehat{\rho }_{R}^{A}$ and $\widehat{\rho }%
_{R}^{B}$ must correspond to physical states that can actually be prepared
in $A$, $B$ considered as stand-alone systems. The sub-system states before
the preparation begins are unimportant, apart from being a separable product
of sub-system states - these might be easily preparable sub-system lowest
energy pure states. This point of view regarding the nature of separable
states is not new, see for example \cite{Bartlett:06, Jones:07a, Masanes:08a}%
. However, other papers focus on the mathematical form for $\hat{\rho}$ and
do not require LOCC preparation for defining what they refer to as separable
but non-local states \cite{Verstraete:03, Schuch:04a}. In the latter papers
the $\widehat{\rho }_{R}^{A}$ and $\widehat{\rho }_{R}^{B}$ are not required
to represent physical states that could be prepared in the isolated
sub-systems. However, for LOCC\ based separable states the joint probability
for measurements of physical quantities $\widehat{\Omega }_{A}$, $\widehat{%
\Omega }_{B}$ for the sub-systems resulting in eigenvalues $\lambda _{Ai}$, $%
\lambda _{Bj}$ will be of the form $P_{A,B}(i,j)=\sum_{R}P_{R}%
\,P_{R}^{A}(i)P_{R}^{B}(j)$, where $\widehat{\Pi }_{Ai}$, $\widehat{\Pi }%
_{Bj}$ are the projectors onto subspaces with eigenvalues $\lambda _{Ai}$, $%
\lambda _{Bj}$, and the factors $P_{R}^{A}(i)=Tr_{A}(\widehat{\Pi }_{Ai}%
\widehat{\rho }_{R}^{A})$, $P_{R}^{B}(j)=Tr_{B}(\widehat{\Pi }_{Bj}\widehat{%
\rho }_{R}^{B})$ give the probabilities that the measurement results $%
\lambda _{Ai}$, $\lambda _{Bj}$ occur if the sub-systems $A$, $B$ are in
states $\widehat{\rho }_{R}^{A}$ and $\widehat{\rho }_{R}^{B}$. If the $%
\widehat{\rho }_{R}^{A}$ and $\widehat{\rho }_{R}^{B}$ did not specify
possible physical states for the sub-systems, then the probabilities $%
P_{R}^{A}(i)$ and $P_{R}^{B}(j)$ would not have a physical meaning. This
property of separable states in which the joint measurement probability is
determined from measurement probabilities of physically possible separate
sub-system states, is the key feature whose absence in non-separable states
led Schrodinger to refer to such states as entangled states. Note that after
preparation, a separable state may evolve into an entangled state, such as
when interactions between the sub-systems take effect.

In addition, for systems of identical bosons we argue that to define the
separable states (and hence the entangled states) one should take into
account the presence of the particle-number superselection rule (SSR) not
only at the global level - as required for any quantum state, but also
crucially at the level of the local subsystems in the case of separable
states. Simply put, the global SSR means that for any quantum state of
massive bosons, $\hat{\rho}$, should commute with the total particle number
in the system $[\hat{\rho},\hat{N}]=0$, and hence no coherences exist
between Fock states\ with differing $N$. For the separable states given by
Eq. (\ref{eq:separable}) for two modes, this means in addition that for
physical sub-system states, $\hat{\rho}_{R}^{X}$ for $X=A,\,B$ should also
commute with the local particle numbers - thus $\hat{\rho}_{R}^{X}$ satisfy $%
[\hat{\rho}_{R}^{X},\hat{N}_{X}]=0$. Note that both $\hat{\rho}$ and the $%
\hat{\rho}_{R}^{X}$ may be mixed states, with statistical mixtures of
particle number states. This additional local particle number SSR
restriction leads to further tests of entanglement based on spin squeezing
which are radically different to some of the well known entanglement
witnesses. As we shall see below, a similar situation applies to other
entanglement tests. In this letter, following an overview of the pertinant
concepts, we will investigate how the spin squeezing and other inequalities
are modified to detect entanglement when local particle number
superselection rules are enforced for separable states.\pagebreak 

\textbf{2. Global and local SSR\bigskip }

The particle-number SSR is a fundamental constraint for systems of massive
bosons. It occurs because physically realizable processes cannot create
states which are coherent superpositions of different Fock states. Thus, all
global operators should commute with the total particle number operator. For
states that are separable, the sub-system states are also required to be
physical states, so analogous constraints must also apply to the
sub-systems, and hence the sub-system states also have no coherences between
Fock states with differing local particle number. This feature is required
for separable states in identical particle systems if - as described above -
joint measurements are to be based on physical probabilities for such
measurements on the sub-systems. Indeed, the SSR also limits the allowed
local operations on the modes (e.g. unitary operations of the form $\widehat{%
U}_{A}\otimes $\ $\widehat{U}_{B}$) -- they too have to commute with the
local particle numbers. Enforcing the local SSR leads to a more restricted
set of separable states than would apply if they are ignored, which can be
expected to modify an entanglement witness that was based upon the
non-restricted definition of separability, in which the $\widehat{\rho }%
_{R}^{A}$ and $\widehat{\rho }_{R}^{B}$ are not required to satisfy the
local particle number SSR. Other authors have applied the SSR at a local
level, see \cite{Verstraete:03, Bartlett:06, Bartlett:07, White:09,
Paterek:11, Kitaev:04a, Vaccaro:08a} for instance. A similar separable state
was used \cite{Goold:09} to show that the visibility of interference fringes
between two spatial modes is an entanglement witness for modes of massive
particles. Non local SSR conforming coherent superpositions of number states
(such as Glauber coherent states) were once thought necessary for describing
interferometry and coherence effects in Bose-Einstein condensates, but it is
now recognised \cite{Leggett:01a, Bach:04a} that valid descriptions of these
effects can be based on Fock states.

We will now make a brief comment about the connection between reference
systems and superselection rules. It has long been known \cite{Aharonov:67}
that by using a suitable reference system one can, at least in principle,
perform protocols (e.g. Ramsay interferometry, dense coding, Bell
inequalities) whose description involves coherences between different
particle numbers \cite{Dowling:06, Heaney:09, Paterek:11}. However, this
apparent inconsistency can be resolved \cite{Bartlett:06b}. The key point is
that the system states involved in a process are described differently by
observers possessing different phase reference systems. If one observer
describes the state as in Eq. (\ref{eq:separable}) in terms of a reference
system where the sub-system density operators $\widehat{\sigma }_{R}^{X}$\
do not conform to the local particle number SSR, then a second observer
without access to this reference system would also describe a separable
state, but now with the transformed sub-system density operators $\hat{\rho}%
_{R}^{X}$\ compatible with the local subsystem SSR. A more detailed
discussion in terms of the $U(1)$\ symmetry group is presented in \cite%
{Bartlett:07} (see also \cite{Dalton:InPrep}, Appendix 4). In this paper, we
work from the viewpoint of the second observer. Phase reference systems such
as Bose-Einstein condensates with large boson numbers - described by the
second observer as being in statistical mixtures of Glauber coherent states
with large fixed amplitudes and all phases having equal weight, and which
satisfies the SSR for the BEC mode - are often involved in protocols such as
Ramsay interferometry, dense coding, Bell inequalities. Bosons from the BEC
reference interact and exchange with the primary system that is involved,
for example two mode \cite{Dowling:06, Heaney:09} or four mode systems \cite%
{Paterek:11}. From the point of view of the first observer, the evolution
from each large amplitude Glauber coherent state of the BEC times the
initital state of the primary system can create states of the primary system
that violate the SSR. However, from the point of view of the second
observer, the evolved state of the primary system conforms to the
super-selection rule following averaging over the BEC phases. In our
approach, separable states are described by an observer with different phase
reference systems with unknown phases for the different sub-systems - as is
appropriate for sub-systems that are separate. On the other hand, entangled
states are described by an observer with a single phase reference system
(albeit one where the phase is unknown), so that entangled states are global
particle number SSR compliant. This is appropriate for composite systems
where the sub-systems are combined into an entangled state on which
measurements for overall system operators can also be performed. In the
present paper the protocol is spin squeezing and we will describe this in
terms of reference systems so that the state Eq.(\ref{eq:separable})
satisfies $[\hat{\rho}_{R}^{X},\hat{N}_{X}]=0$\ for $X=A,\,B$, which can be
achieved in ultra-cold gas experiments.\bigskip 

\textbf{3. Spin squeezing requires entanglement\bigskip }

Under the global and local particle-number superselection rule requirements
for physical states of indistinguishable particles, the spin squeezing
inequalities Eq. (\ref{eq:spinsq1}) are satisfied only if entanglement is
present between the modes $A$ and $B$. The proof is as follows. Using
realization, $\hat{\rho}_{R}$, of the separable state Eq. (\ref{eq:separable}%
), the operators in the variance are $\langle \hat{S}_{x}^{2}\rangle
_{R}=\langle \hat{S}_{y}^{2}\rangle _{R}=1/4\langle \hat{N}\rangle
_{R}+1/2n_{R}^{A}n_{R}^{B}$ and $\langle \hat{S}_{x}\rangle _{R}=\langle 
\hat{S}_{y}\rangle _{R}=0$, with $n_{R}^{X}=\left\langle \hat{N}%
_{X}\right\rangle _{R}$. Then, since for any operator $\hat{\Omega}$ the
inequality $\langle \Delta \hat{\Omega}^{2}\rangle \geq \sum_{R}P_{R}\langle
\Delta \hat{\Omega}^{2}\rangle _{R}$ applies \cite{Hofmann:03a}, the
inequality $\langle \Delta \hat{S}_{x}^{2}\rangle \geq \sum_{R}P_{R}\langle
\Delta S_{x}^{2}\rangle _{R}=\sum_{R}P_{R}(\frac{1}{4}\langle \hat{N}\rangle
_{R}+\frac{1}{2}n_{R}^{A}n_{R}^{B})$ holds and likewise for $\langle \Delta 
\hat{S}_{y}^{2}\rangle $. For SSR-restricted separable states, Eq. (\ref%
{eq:separable}), the expectation values of the z spin component $\frac{1}{2}%
\langle \hat{S}_{z}\rangle =\sum_{R}P_{R}\frac{1}{4}(n_{R}^{B}-n_{R}^{A})$
can be bounded from above as $\frac{1}{2}|\langle \hat{S}_{z}\rangle |\leq
\sum_{R}P_{R}\frac{1}{4}|n_{R}^{B}-n_{R}^{A}|\leq \sum_{R}P_{R}\frac{1}{4}%
|n_{R}^{B}+n_{R}^{A}|$. Thus, $\langle \Delta \hat{S}_{x}^{2}\rangle -\frac{1%
}{2}|\langle \hat{S}_{z}\rangle |\geq \sum_{R}P_{R}\frac{1}{2}%
(n_{R}^{A}n_{R}^{B})\geq 0$ and likewise $\langle \Delta \hat{S}%
_{y}^{2}\rangle -\frac{1}{2}|\langle \hat{S}_{z}\rangle |\geq 0$. This shows
for general separable states in Eq.(\ref{eq:separable}) over two modes, $A$
and $B$ in which the local mode SSR\ applies, the spin variances in both the 
$x$ and $y$ directions satisfy 
\begin{equation}
\langle \Delta \hat{S}_{x}^{2}\rangle \geq \frac{1}{2}|\langle \hat{S}%
_{z}\rangle |\qquad \langle \Delta \hat{S}_{y}^{2}\rangle \geq \frac{1}{2}%
|\langle \hat{S}_{z}\rangle |  \label{Eq.SpinSqResult}
\end{equation}%
and hence there is no spin squeezing of $\hat{S}_{x}$ with respect to $\hat{S%
}_{y}$ or vice versa \ It is straightforward to show that analogous results
apply to the other pairs of spin components $\widehat{S}_{y},\widehat{S}_{z}$%
\ and $\widehat{S}_{z},\widehat{S}_{x}$\ or to each spin component and any
perpendicular spin component. Thus we have now shown that spin squeezing in
any spin component is a test of entanglement -entanglement being defined
here in terms of separable states satisfying the local particle number SSR.
There are, of course, entangled states which do not show spin squeezing. One
example is the famous Schrodinger cat like $NOON$\ state, $(\left\vert
N,0\right\rangle +\left\vert 0,N\right\rangle )/\sqrt{2}$, which describes
all the particles in mode $A$\ and none in mode $B$\ superposed with no
particles in mode $A$\ and all particles in mode $B$. For large $N$ such
states are however notoriously difficult to create in experiments. The $NOON$
state is global particle number SSR compliant, and its density operator is
the same for any observer with a different overall phase reference system.
Note that for an observer with separate unrelated phase reference systems
for the two modes, the state is seen as a separable state, whose density
operator is a 50:50 mixture with one state having $N$ bosons in mode $A$ and
none in mode $B$, the other having no bosons in mode $A$ and $N$ in mode $B$%
. \pagebreak 

\textbf{4. Other spin squeezing tests and local SSR separable states\bigskip 
}

There are several entanglement tests, based both on spin squeezing and other
protocols that can be used to detect entanglement in systems, where the
definition of entanglement is based on separable states which do not satisfy
the physically based particle-number superselection rule. Of course any test
that is derived for arbitrary $\hat{\rho}_{R}^{X}$\ must also demonstrate
the present entanglement based on $[\hat{\rho}_{R}^{X},\hat{N}_{X}]=0$.
Several such tests are discussed elsewhere \cite{Dalton:InPrep}. However, it
is interesting and necessary for experiments to consider whether additional
tests arise when they are re-derived for systems of identical massive bosons
based on the physical definition of a separable state (\ref{eq:separable}),
where the local SSR is satisfied. We will focus on certain commonly used
inequalities in this paper, although other tests also exist (see \cite{Toth,
Duan}).\medskip 

\textit{4.1} \textit{Hillery et al 2006 spin squeezing test\medskip }

A paper by Hillery and co-workers \cite{Hillery:06} derives an entanglement
witness (Eq.(\ref{Eq.Hillery06})) based upon spin squeezing, based on
general (non-SSR-restricted) separable states of the EM field. It has been
used to detect entanglement in e.g. the following proposals for trapped
ultra-cold gases \cite{He:12a, He:12b}. We confirm that the main result of 
\cite{Hillery:06} is still valid for the SSR restricted state (\ref%
{eq:separable}), but provide an example where the Hillery test based on the
sum of $\widehat{J}_{x},\widehat{J}_{y}$\ variances for rotated spin
components fails to detect entanglement in the relevant modes, whereas a
simple spin squeezing test (as in Eqs.(\ref{Eq.SpinSqResult})) involving a
rotated component $\widehat{J}_{y}$\ is satisfied.

Hillery et al \cite{Hillery:06} show that for general separable states 
\begin{equation}
\langle \Delta S_{x}^{2}\rangle +\langle \Delta S_{y}^{2}\rangle \geq \frac{1%
}{2}\langle \hat{N}\rangle ,  \label{Eq.Hillery06}
\end{equation}%
To obtain this result it is found that for a product state $\hat{\rho}_{R}$\
that $\langle \Delta S_{x}^{2}\rangle _{R}+\langle \Delta S_{y}^{2}\rangle
_{R}=(\frac{1}{2}\langle \hat{N}\rangle _{R}+n_{R}^{A}n_{R}^{B}-|\langle 
\hat{a}\rangle _{R}|^{2}|\langle \widehat{b}^{\dag }\rangle _{R}|^{2})$,
where for non SSR compliant $\hat{\rho}_{R}^{X}$\ the terms $\langle \hat{a}%
\rangle _{R}$\ and $\langle \widehat{b}^{\dag }\rangle _{R}$\ are not
necessarily zero. Noting that $|\langle \hat{a}\rangle |^{2}\leq \langle 
\hat{N}_{a}\rangle $\ and likewise for mode $b$, it follows that $\langle
\Delta S_{x}^{2}\rangle _{R}+\langle \Delta S_{y}^{2}\rangle _{R}\geq \frac{1%
}{2}(\langle \hat{N}\rangle _{R}$. From $\langle \Delta \hat{\Omega}%
^{2}\rangle \geq \sum_{R}P_{R}\langle \Delta \hat{\Omega}^{2}\rangle _{R}$\
the result in Eq.(\ref{Eq.Hillery06}) then follows. For SSR-restricted
separable states however, the result $\langle \Delta S_{x}^{2}\rangle
_{R}+\langle \Delta S_{y}^{2}\rangle _{R}\geq \frac{1}{2}(\langle \hat{N}%
\rangle _{R}$\ still holds because $n_{R}^{A}n_{R}^{B}\geq 0$\ alone is
needed. The Hillery et al spin squeezing test for entanglement then is $%
\langle \Delta S_{x}^{2}\rangle +\langle \Delta S_{y}^{2}\rangle <\frac{1}{2}%
\langle \hat{N}\rangle $. If this applies then the state cannot be
separable. 

Note that we have previously shown that for local SSR compliant separable
states that $\langle \Delta S_{x}^{2}\rangle +\langle \Delta
S_{y}^{2}\rangle \geq |\langle \hat{S}_{z}\rangle |$. The quantity $|\langle 
\hat{S}_{z}\rangle |$\ is smaller than $\frac{1}{2}\langle \hat{N}\rangle $\
since $|\langle \hat{S}_{z}\rangle |\leq \sum_{R}P_{R}\frac{1}{2}%
|n_{R}^{B}-n_{R}^{A}|\leq \sum_{R}P_{R}\frac{1}{2}|n_{R}^{B}+n_{R}^{A}|=%
\frac{1}{2}\langle \hat{N}\rangle $. However, $\langle \Delta
S_{x}^{2}\rangle +\langle \Delta S_{y}^{2}\rangle <|\langle \hat{S}%
_{z}\rangle |$\ is not a\ valid entanglement test because there are no
quantum states where this is true, as noted in \cite{Hillery:06}.

Although it may be thought that because $|\langle \hat{S}_{z}\rangle |$\ is
smaller than $\frac{1}{2}\langle \hat{N}\rangle $\ the Hillery test
involving $\langle \Delta S_{x}^{2}\rangle +\langle \Delta S_{y}^{2}\rangle $%
\ $<$\ $\frac{1}{2}\langle \hat{N}\rangle $\ would be more likely to
demonstrate entanglement in the modes $A$, $B$\ than the tests obtained in
this paper such as showing $\langle \Delta \hat{S}_{x}^{2}\rangle <\frac{1}{2%
}|\langle \hat{S}_{z}\rangle |$\ or $\langle \Delta \hat{S}_{y}^{2}\rangle <%
\frac{1}{2}|\langle \hat{S}_{z}\rangle |$, this is not always the case. The
relative phase eigenstate \cite{Barnett:89a, Dalton:12a} is an entangled
pure state for $N$ bosons defined by $\left\vert \frac{N}{2},\theta
_{p}\right\rangle =(\sum_{k=-N/2}^{k=+N/2}\exp (ik\theta _{p})\,\left\vert
N/2-k\right\rangle ^{A}\otimes \left\vert N/2+k\right\rangle ^{B})/\sqrt{N+1}
$, where $\theta _{p}=p(2\pi /(N+1))$ with $p=-N/2,-N/2+1,..,+N/2$ specifies
the relative phase. In terms of spin components $\widehat{S}_{x},\widehat{S}%
_{y},\widehat{S}_{z}$ the covariance matrix \cite{Jaaskelainen:06a} for spin
fluctuations is non-diagonal and spin squeezing does not occur. As the
variances are such that $\langle \Delta \hat{S}_{x}^{2}\rangle +\langle
\Delta \hat{S}_{y}^{2}\rangle =(\frac{1}{6}-\frac{\pi ^{2}}{64})N^{2}$ and
which exceeds $\frac{1}{2}N$, the Hillery test for entanglement also fails
even though the state is entangled for modes $A$, $B$.\ Principal spin
components $\widehat{J}_{x}=\widehat{S}_{z},\widehat{J}_{y}=(\sin \theta
_{p})\,\widehat{S}_{x}+(\cos \theta _{p})\,\widehat{S}_{y},\widehat{J}%
_{z}=-(\cos \theta _{p})\,\widehat{S}_{x}+(\sin \theta _{p})\,\widehat{S}_{y}
$ are obtained by a rotation, and for these the covariance matrix is
diagonal with $\langle \Delta \widehat{J}_{x}^{2}\rangle =\frac{1}{12}%
N^{2},\langle \Delta \widehat{J}_{y}^{2}\rangle =\frac{1}{4}+\frac{1}{8}\ln N
$ and $\langle \Delta \widehat{J}_{z}^{2}\rangle =(\frac{1}{6}-\frac{\pi ^{2}%
}{64})N^{2}$ for large $N$. The mean values are $\langle \widehat{J}%
_{x}\rangle =\langle \widehat{J}_{y}\rangle =0,\langle \widehat{J}%
_{z}\rangle =-\frac{\pi }{8}N$ (see \cite{Dalton:12a} for details). The
principal spin components are related to annihilation operators $\widehat{c},%
\widehat{d}$ for new modes $C,D$ via expressions of the form $\widehat{J}%
_{x}=(\widehat{d}^{\dag }\widehat{c}+\widehat{c}^{\dag }\widehat{d})/2$,
etc, where $\widehat{a}=-\exp (\frac{1}{2}i\theta _{p})(\widehat{c}-\widehat{%
d})/\sqrt{2}$ and $\widehat{b}=-\exp (-\frac{1}{2}i\theta _{p})(\widehat{c}+%
\widehat{d})/\sqrt{2}$, and the relative phase state can be rewritten as a
linear combination of Fock states for the new modes $\left\vert
N/2-l\right\rangle ^{C}\otimes \left\vert N/2+l\right\rangle ^{D}$ with $%
l=-N/2,.,+N/2$. Note that there must be terms with differing $l$\ since $%
\widehat{J}_{z}(\left\vert N/2-l\right\rangle ^{C}\otimes \left\vert
N/2+l\right\rangle ^{D})=l(\left\vert N/2-l\right\rangle ^{C}\otimes
\left\vert N/2+l\right\rangle ^{D})$\ and $\langle \Delta \widehat{J}%
_{z}^{2}\rangle >0$. The relative phase state is therefore an entangled
state for the modes $C,D$. In the relative phase state $\widehat{J}_{y}\ $is
squeezed with respect to $\widehat{J}_{x}$ so the spin squeezing test for
entanglement of modes $C,D$ based on separable states consistent with the
local particle number SSR is satisfied. However, for the Hillery test $%
\langle \Delta \widehat{J}_{x}^{2}\rangle +\langle \Delta \widehat{J}%
_{y}^{2}\rangle \approx \frac{1}{12}N^{2}$, which considerably exceeds $%
\frac{1}{2}N$. Thus the Hillery test for entanglement fails.\medskip 

\textit{4.2} \textit{Sorensen et al 2001 spin squeezing test\medskip }

Care must be taken when applying a spin squeezing entanglement based witness
derived in an early paper by Sorensen et al \cite{Sorensen:01} for systems
of $N$\ distinguishable two-level particles to the situation when the
particles are identical \cite{Hyllus:12a}. The witness shows that
entanglement exists when the state satisfies the following spin squeezing
inequality 
\begin{equation}
\xi ^{2}\equiv \frac{N(\Delta \widehat{S}_{z})^{2}}{\langle \widehat{S}%
_{x}\rangle ^{2}+\langle \widehat{S}_{y}\rangle ^{2}}<1.  \label{eq:sorensen}
\end{equation}%
The proof is based on writing the separable state density operator in the
form $\hat{\rho}_{sep}=\sum_{R}P_{R}\,\hat{\rho}_{R}^{1}\otimes \hat{\rho}%
_{R}^{2}$\ $\otimes \hat{\rho}_{R}^{3}...$where $\hat{\rho}_{R}^{i}$\ is the
density operator for the $i$th distinguishable particle, whose internal
states are $\left\vert \phi _{a}(i)\right\rangle $ and $\left\vert \phi
_{b}(i)\right\rangle $. The density matrix for $\hat{\rho}_{R}^{i}$\ is a $%
2\times 2$\ matrix and the spin operators are defined by expressions such as 
$\widehat{S}_{x}=\sum_{i}(\left\vert \phi _{b}(i)\right\rangle \left\langle
\phi _{a}(i)\right\vert +\left\vert \phi _{a}(i)\right\rangle \left\langle
\phi _{b}(i)\right\vert )/2$\ etc., which are sums over the particles. The
hermiticity, positiveness, unit trace for all the density matrices lead to $%
\xi ^{2}\geq 1$\ for these separable states of two level distinguishable
particles. This result cannot just be applied to identical bosonic particles
without further development, since the Sorensen separable state density
operator does not satisfy the symmetrization principle. Also Benatti et al 
\cite{Benatti:11} have shown that this inequality diverges for two mode
states where the local particle-number superselection rule is enforced, as
is easily seen by noting that $\langle \widehat{S}_{x}\rangle =\langle 
\widehat{S}_{y}\rangle =0$\ for a physical separable state, Eq.(\ref%
{eq:separable}) involving single modes as subsystems. \ Just considering $N$
boson states for a total of two modes is not adequate. One way to revise the
Sorensen et al result to an equivalent theory now based on mode entanglement
involves sub-systems each consisting of two modes, which we may list as $%
\alpha k$, where $\alpha =a,b$\ and $k$\ lists modes with the same $\alpha $%
. The $k$\ may correspond to spatial modes localised on different lattice
sites. For identical particles occupying these modes the spin operators

$\widehat{S}_{x}=\sum_{k}\sum_{i}$\ $(\left\vert \phi _{bk}(i)\right\rangle
\left\langle \phi _{ak}(i)\right\vert +\left\vert \phi _{ak}(i)\right\rangle
\left\langle \phi _{bk}(i)\right\vert )/2$\ etc become Schwinger spin
operators $\widehat{S}_{x}=\sum_{k}(\hat{b}_{k}^{\dag }\hat{a}_{k}+\hat{a}%
_{k}^{\dag }\hat{b}_{k})/2$\ $=\sum_{k}\widehat{S}_{x}^{k}$ etc, where the $%
\widehat{S}_{\gamma }^{k}$\ and $\widehat{S}_{\gamma }$satisfy the usual
commutation rules. The sub-system density operators $\hat{\rho}_{R}^{k}$\
now refer to two mode sub-systems and can be made consistent with the local
particle number SSR by requiring that $[\hat{\rho}_{R}^{k},\widehat{N}%
_{a}^{k}+\widehat{N}_{b}^{k}]=0$. The symmetrization principle automatically
applies for $\hat{\rho}_{sep}=\sum_{R}P_{R}\,\tprod\limits_{k}(\hat{\rho}%
_{R}^{k})^{\otimes }$ in this second quantization treatment. If in addition,
we require that the $\hat{\rho}_{R}^{(k)}$ are density operators for a
single boson, then the bosons in different $k$\ modes are effectively
distinguishable. The proof in Sorensen et al \cite{Sorensen:01} then applies,%
\textbf{\ }noting that for two mode sub-systems $\left\langle \widehat{S}%
_{\gamma }^{k}\right\rangle _{R}$\ and hence $\left\langle \widehat{S}%
_{\gamma }\right\rangle $\ $(\gamma =x,y,z)$\ are not necessarily zero.
States of this type are routinely created in experiments, for instance in 
\cite{Mandel:03}, where the gas is frozen deep in the Mott regime with unity
filling and each atom has two accessible internal states. The sites then act
like distinguishable qubits. The physicality of a separable state is thus
guaranteed and there may be entanglement between the modes within each
subsystem. A similar proof extending the test of Sorensen et al to identical
two level systems is given by Hyllus et al \cite{Hyllus:12a} based on a
particle entanglement approach. In their approach bosons in different
external modes (such as the different $k$) are treated as distinguishable
and the symmetrization principle is ignored for such bosons.

It should be noted that for separable states where the $\hat{\rho}_{R}^{X}$\
are required to satisfy further conditions in addition to the local particle
number SSR , the entanglement tests will differ from those where the
additional conditions are absent. The requirement that the sub-system
density operators $\hat{\rho}_{R}^{k}$\ are restricted to one boson states
is an example of such an additional condition.\bigskip 

\textbf{5. Non spin squeezing tests and local SSR separable states}\bigskip

In \cite{Hillery:06, Hillery:09} it was found that for separable states
based on arbitrary sub-system density operators $\hat{\rho}_{R}^{X}$\ 
\begin{equation}
|\langle \hat{a}^{m}(\hat{b}^{\dag })^{n}\rangle |^{2}\leq \langle (\hat{a}%
^{\dag })^{m}\hat{a}^{m}(\hat{b}^{\dag })^{n}\hat{b}^{n}\rangle \mathbf{\ }
\label{Eq.Hilller09Inequality}
\end{equation}%
for $m,n=0,1,..$so that $|\langle \hat{a}^{m}(\hat{b}^{\dag })^{n}\rangle
|^{2}>\langle (\hat{a}^{\dag })^{m}\hat{a}^{m}(\hat{b}^{\dag })^{n}\hat{b}%
^{n}\rangle $\ is a test for entanglement. This test must also apply for
separable states where the local SSR is satisfied. A particular case of the
test is $|\langle \hat{a}\hat{b}^{\dag }\rangle |^{2}=|\langle \hat{a}^{\dag
}\hat{b}\rangle |^{2}>\langle \hat{N}_{a}\hat{N}_{b}\rangle $\ \ 

However, for separable states satisfying the local particle number SSR we
can easily show that 
\begin{equation}
|\langle \hat{a}^{m}(\hat{b}^{\dag })^{n}\rangle |^{2}=0.
\label{Eq.SSRInequality}
\end{equation}%
Hence $|\langle \hat{a}^{m}(\hat{b}^{\dag })^{n}\rangle |^{2}>0$\ is a test
for entanglement based on local SSR compliant separable states. Since $%
|\langle \hat{a}^{m}(\hat{b}^{\dag })^{n}\rangle |^{2}$\ is merely required
to be non-zero this test is easier to satisfy than the Hillery one based on
Eq.(\ref{Eq.Hilller09Inequality}). A particular case of the test is $%
|\langle \hat{a}\hat{b}^{\dag }\rangle |^{2}>0$. The last result also
follows from $\langle \widehat{S}_{x}\rangle =\langle \widehat{S}_{y}\rangle
=0$ for SSR compatible states. Hence this is a simpler test for entanglement
than $|\langle \hat{a}\hat{b}^{\dag }\rangle |^{2}>\langle \hat{N}_{a}\hat{N}%
_{b}\rangle $\ and has been used instead to detect entanglement \cite%
{Goold:09}\bigskip .

\textbf{6. Experimental considerations}\bigskip

A recent experiment \cite{Gross:10}, uses the inequality (\ref{eq:sorensen})
to detect the entanglement in an ultra-cold gas. Here, despite their
indistinguishability, they consider the particles themselves as the
sub-systems and generate `entanglement' via their internal degrees of
freedom. Due to this, the inequality (\ref{eq:sorensen}) would be valid as
described above if distinguishability could be recovered. This while
technically very difficult, can in principle be achieved. For instance, the
gas could be frozen (without disturbing the internal states of the particles
- into the Mott phase of an optical lattice so that the system behaves as a
set of distinguishable qubits. Such a scheme has been considered above.

On the otherhand, an earlier experiment \cite{Esteve:08} considers the
entanglement of two spatial field modes. Here the application of inequality (%
\ref{eq:sorensen}) should be reconsidered due to the fact that the
subsystems are single modes (with no internal structure).

The experiments only involve a single test for entanglement, and it would be
desirable to confirm entanglement via an independent test.\bigskip

\textbf{7. Conclusions}\bigskip

Only three ground breaking experiments \cite{Gross:10, Esteve:08,
experiments} have shown spin squeezing in cold atom systems, from which the
presence of entanglement is inferred. As such the field is still very much
in its infancy. Our results will advance the field by allowing a greater
understanding of the role of indistinguishability when detecting
entanglement. A more extensive presentation of the work in this paper is in
preparation \cite{Dalton:InPrep}.\bigskip

\textbf{Acknowledgements}\bigskip

The authors thank S M Barnett, J F Corney, P D Drummond, J Jeffers, K
Molmer, D Oi, M D Reid, K Rzazewski, T Rudolph, J A Vaccaro and V Vedral for
helpful discussions. BJD thanks the Science Foundation of Ireland for
funding this research via an E\ T S Walton Visiting Fellowship.\bigskip


\end{document}